\documentstyle[aps,epsf]{revtex}  

\begin{document}        

\preprint{
\vbox{\hbox{JHU--TIPAC--99001} }}

\baselineskip 14pt
\title{Rescattering Effects in Heavy Quark Decays}
\author{Alexey A. Petrov}
\address{Department of Physics and Astronomy,
Johns Hopkins University,
Baltimore, Maryland 21218}
%
\maketitle              

\begin{abstract}        
 We review some recent developments in the studies of direct 
$CP$ violation and final state interactions in weak decays of heavy quarks. 
\
\end{abstract}   	

\section{Introduction and Motivation}

Heavy quark decays serve as a powerful tool for testing the Standard 
Model and, since it probes the quarks of all three generations, provide 
invaluable possibilities to study CP violation.  However, the
interpretation of experimental observables in terms of fundamental parameters
is often less than clear. Rare hadronic decays of $B$ mesons, for
example, proceed through both tree level Cabibbo-suppressed amplitudes and
through one loop penguin amplitudes. On the one hand, this situation allows direct
CP violating effects. On the other, these contributions complicate
the extraction of Cabibbo-Kobayashi-Maskawa (CKM) angles and, in particular,
the angle $\gamma\equiv\arg\left[-{V_{ud}V_{ub}^*/V_{cd}V_{cb}^*}\right].$

The decay of a heavy hadron produces quarks in the
final state. Because of their strong QCD interactions they
continue to interact after the weak transition took place.
Even after they have formed hadrons, there are still strong forces
between them, and therefore, the problem of the
{\em final state rescatterings or final state interactions (FSI) } is an 
important part of the physics of nonleptonic $b$-decays.
The most pronounced effect of FSI is clearly in direct
CP-violation where one compares the rates of a $B$-meson
decay with the charged conjugated process.
The corresponding asymmetries between the two
decays depend on both a weak (CKM) and a strong
rescattering phase provided by the FSI. Furthermore, a 
non-vanishing asymmetery requires two different final 
states produced by different weak amplitudes
which can go into each other by a strong interaction rescattering.
Thus, FSI directly affect the asymmetries and their size can be interpreted
in terms of fundamental parameters {\it only} if these FSI  phases are calculable.

As an example, let us first look at the recently measured combined
branching ratios for $B^\pm\to\pi^\pm K$ and $B_d \to \pi^\mp K^\pm $~\cite{fknp}.  
In the Standard Model,
these decays are mediated by the $\Delta B = 1$ Hamiltonian, which takes the form
\begin{eqnarray} \label{HeffCP} 
{\cal H}_{\rm eff} = \frac{G_F}{\sqrt{2}}\Bigl[
 V_{cb} V_{cs}^* \left(\sum_{i=1}^{2}
C_i Q_i^{c s} + \sum_{i = 3}^6 C_i Q_i^s + \sum_{i = 7}^{10}
C_i Q^s_i \right)
+ V_{ub} V_{us}^* \left(\sum_{i=1}^{2}
C_i Q_i^{u s} + \sum_{i = 3}^6 C_i Q_i^s + \sum_{i = 7}^{10}
C_i Q^s_i \right)\Bigr] + {\rm H.c.},
\end{eqnarray}
The flavor structures of the current-current, QCD penguin, and electroweak
penguin (EWP) operators are, respectively,
$Q_{1,2}^{qs} \sim \bar s q  \bar q b $,
$Q_{3,..,6}^{s} \sim \bar s b \sum \bar q' q' $, and
$Q_{7,..,10}^{s} \sim \bar s b \sum e_{q'} \bar q' q' $, the sum is over
light quark flavors. The Wilson coefficients $C_i$ are renormalization scale dependent,
$C_{1,2} (1~GeV) = {\cal O}(1)$, $C_{3,..,6,9}(1~GeV) = {\cal O}(10^{-2})$, and
$C_{7,8,10}(1~GeV) \le {\cal O}(10^{-3} )$. Let us expand the decay amplitudes of interest 
according to their dependence on the elements of the CKM matrix,
\begin{eqnarray} \label{Camps}
A(B^+ \to \pi^+ K^0) = A_{cs}^+  - A_{us}^+ e^{i \gamma} e^{i \delta_+},~~~~~
A(B^- \to \pi^- \bar K^0) = A_{cs}^+ - A_{us}^+ e^{-i \gamma} e^{i \delta_+},
\nonumber\\
A(B^0 \to \pi^- K^+) =  A_{cs}^0 - A_{us}^0 e^{i \gamma} e^{i
\delta_0},~~~~~ A(\bar B^0 \to \pi^+ K^-) =   A_{cs}^0
- A_{us}^0 e^{-i \gamma} e^{i \delta_0},
\end{eqnarray}
where $\delta_0$ and $\delta_+ $ are CP-conserving phases induced by the strong
interaction. The first and second terms in each amplitude correspond to matrix
elements of the  first and second terms in ${\cal H}_{eff}$ (or their Hermitian
conjugates), respectively. Note that each term is by itself scheme and
renormalization scale  independent.

One can asses the
expected relative contributions of the operators in ${\cal H}_{\rm eff}$ to a given
exclusive decay mode.  The electroweak penguin operators are commonly neglected,
since the contributions with a sizable Wilson coefficient, $C_9Q^s_9$, are color
suppressed or require rescattering from intermediate states.  In this case isospin
symmetry of the strong interactions leads to the  simplification
$A_{cs}^0 =  A_{cs}^+ $.
It is now believed that the current-current operator
contributions to $A_{cs}^{0,+}$ are roughly of same order as the
QCD penguin operator contributions.
The contribution of the current-current operators to
$A_{us}^0$ is also expected to be of the same order, despite the CKM
suppression,
because of the large value of $C_2 $, namely, $ V_{ub} V^*_{us}\, C_2
\sim V_{cb} V_{cs}^*\, C_{3,..6} $.
However, since for $B^\pm \to \pi^\pm K $ the relevant quark transition is
$b \to d \bar d s$,  one might expect the size of
$A_{us}^+ $ relative
to $A_{cs}^+ $ to be highly suppressed by the
small ratio $|V_{ub}V_{us}^* /V_{cb}V_{cs}^* |\sim 0.02$.  This would hold equally
for the current-current and penguin operators. If, indeed, $r_+ = A_{us}^+
/A_{cs}^+ \sim |V_{ub}V_{us}^* / V_{cb}V_{cs}^* |$  is a good approximation, then
there are two important consequences:

$(i)$ Direct CP violation could be observed, in principle, through
the CP asymmetry $\mbox{${\cal A}^{\rm dir}_{\rm CP}$}\equiv\mbox{${\cal A}^{\rm dir}_{\rm CP}$}(B^+\to\pi^+K^0)$,
\begin{equation}\label{ABpiK}
\mbox{${\cal A}^{\rm dir}_{\rm CP}$}={
BR(B^+\to\pi^+K^0)-BR(B^-\to\pi^-\bar K^0)\over
BR(B^+\to\pi^+K^0)+BR(B^-\to\pi^-\bar K^0)}
={2r_+\sin\gamma\sin\delta_+\over1-2r_+\cos\gamma\cos\delta_++r_+^2}\,.
\end{equation}
However, it would be small,
$
\mbox{${\cal A}^{\rm dir}_{\rm CP}$}(B^+\to\pi^+ K^0)\leq{\cal O}(\lambda^2),
$
where $\lambda \simeq0.22 $ is the Wolfenstein parameter.
``Hard'' FSI estimates, where the $u$ quarks in $Q_{1,2}^{us} $ are 
treated as a perturbative loop, give
$\mbox{${\cal A}^{\rm dir}_{\rm CP}$} \sim 1\% $.

$(ii)$ Model-independent bounds could be
obtained for the angle $\gamma$ using only the {\it combined\/}
branching ratios  $BR(B^\pm \to \pi^\pm K)$ and
$BR(B_d \to \pi^\mp K^\pm)$ \cite{fm,GNPS}.
One can construct the ratio
\begin{equation} \label{ratio}
R = \frac{BR(B^0 \to \pi^- K^+) + BR(\bar B^0 \to \pi^+ K^-)}{
BR(B^+ \to \pi^+ K^0) + BR(B^- \to \pi^- \overline{K^0})}=
\left({A_{cs}^0\over A_{cs}^+}\right)^2{1-2r_0\cos\gamma\cos\delta_0+r_0^2
\over1-2r_+\cos\gamma\cos\delta_++r_+^2}\,,
\end{equation}
where $r_0=A_{us}^0 /A_{cs}^0$.
If $A_{us}^+ $ and EWP operator contributions are negligible,
the ratio~(\ref{ratio}) takes the simple form
\begin{equation} \label{theory}
R = 1 - 2 r_0 \cos \gamma \cos \delta_0 + r_0^2\,.
\end{equation}
The observable $R$ may be minimized with respect to the
parameter $r_0$, which (as $\cos^2\delta_0\le1$) leads 
to the bound
\begin{equation} \label{bound}
\sin^2 \gamma  \leq R.
\end{equation}
If true, a stringent bound on $\gamma$ would be obtained if the
reported by CLEO value $R_{\rm exp} = 1.00 \pm 0.40$~\cite{CLEObr}
turns out to be smaller than unity within experimental 
errors~\cite{rosnergronau}.

Rare $B$ decays, like $B\to\pi K$, are suppressed in the Standard Model by
either CKM matrix elements or small Wilson coefficients. Thus, these
decays are potentially sensitive to New Physics. In the
presence of New Physics, a large CP asymmetry can be induced,
${\cal A}^{\rm dir}_{\rm CP} \gg 1\% $
and $R$ can be modified to violate the bound (\ref{bound}).
The analysis leading to (\ref{bound}), however, explicitly assumes that the 
CKM angle $\gamma$ does not enter the theoretical expression for the charged
decay amplitudes (\ref{Camps}), i.e. the absence of large contributions from 
the operators $Q_{1,2}^{us}$. This assumption is
based on the observation that the quark level decay $b\to d\bar ds$ is not mediated
direcly by $Q_{1,2}^{us}$.  However, this treatment of the dynamics ignores
the effects of soft rescattering effects at long distances, which can include the
exchange of global quantum numbers such as charge and strangeness.
In the absence of an argument that parton-hadron duality should hold in
exclusive processes involving pions and kaons,
one must conclude that the long distance physics of meson rescattering is not
probed by the analysis of the final state rescattering based on perturbative QCD. 
In addition, the rescattering in question
is {\it inelastic,} despite its quasi-elastic kinematics, and cannot be studied
adequately in any model of purely {\it elastic\/} final state phases.
Let us proceed with our example and investigate the impact of final state rescattering 
on the CP asymmetry $\mbox{${\cal A}^{\rm dir}_{\rm CP}$}(B^\pm\to\pi^\pm K)$ 
and the ratio $R$. The rescattering process involves an intermediate on-shell 
state $X$, such that $B\to X\to K\pi$.  In particular, we assume that there 
exists a generic (multibody) state $K n\pi$. The charged and neutral channel 
amplitudes can then be written as
\begin{eqnarray} \label{multibody}
A(B^+ \to K n\pi) &=& A_{cs}^{n+}  - A_{us}^{n+} e^{i \gamma}
e^{i \delta_+^n}\,,\\ \nonumber
A(B^0 \to K n\pi) &=& A_{cs}^{n0} - A_{us}^{n0} e^{i \gamma}
e^{i \delta_0^n}\,.
\end{eqnarray}
Rescattering contributions, again
decomposed according to their dependence on CKM factors,
are given by
\begin{eqnarray} \label{amplitudes}
A(B^+ \to K n\pi \to \pi^+ K^0) &=&
S_1^n A_{cs}^{n+} -
S_2^n A_{us}^{n+} e^{i \gamma}\,,\nonumber\\
A(B^0 \to K n\pi \to \pi^- K^+) &=&
S_3^n A_{cs}^{n0}
- S_4^n A_{us}^{n0} e^{i \gamma}\,,
\end{eqnarray}
where $S_i^n$ is the complex amplitude for rescattering from
a given multibody final state to the channel of interest.
In the limit of isospin symmetry $A_{cs}^+=A_{cs}^0$, and this equality is
not spoiled by rescattering effects.  The $i=1,3,4$ rescattering amplitudes
can be absorbed into the unknown amplitudes in Eq.~(\ref{Camps}).

Let us assume that the rescatterings of transitions mediated
by $Q_{1,2}^{us}$ are significant enough to dominate $A^+_{us}$, so 
$A_{us}^+ e^{i\delta_+} = \sum_n S_2^n A_{us}^{n+}$,
and define $\epsilon=A_{us}^+/A_{cs}^+$.  Let us also assume that rescattering effects
do not dominate the overall decay, so we may retain just terms linear in
$\epsilon$. Then,
$\mbox{${\cal A}^{\rm dir}_{\rm CP}$}$ of Eq.~(\ref{ABpiK}) and
$R$ of Eq.~(\ref{ratio}) take the form
\begin{equation}\label{corrACP} \label{corrR}
\mbox{${\cal A}^{\rm dir}_{\rm CP}$}
={2 \epsilon \sin\gamma\sin\delta_+
\over1-2 \epsilon \cos\gamma\cos\delta_+}\,, ~~
R={1  - 2r_0 \cos \gamma \cos \delta_0 + r_0^2\over
1 - 2 \epsilon \cos \gamma \cos \delta_+ }\,.
\end{equation}
Once again, we may extremize $R$ with respect to the unknown $r_0$,
\begin{equation} \label{newbound1}
R \geq \frac{1 - \cos^2 \gamma \cos^2 \delta}
{1 - 2 \epsilon \cos \gamma \cos \delta_+}\,.
\end{equation}
Using the same arguments as before with respect to the strong
phases $\delta_0$ and $\delta_+$, we find the new bound
\begin{equation}\label{FMmod}
  \sin^2\gamma\le R(1+2\epsilon\sqrt{1-R}),
~~
\mbox{or} ~~
|\cos \gamma|\geq \sqrt{1-R}-\epsilon R\,.
\end{equation}
It is clear that
even a small rescattering amplitude $\epsilon \sim 0.1$ could induce
a significant shift in the bound on $\gamma$ deduced from $R$.
For example, the fractional correction to the bound on
$|\cos\gamma|$ is $\Delta\equiv \epsilon R/\sqrt{1-R}$.  The value of $\Delta$ is a
strong function of the experimentally observed $R_{\rm exp}$,
$\Delta\simeq \epsilon ~\mbox{for}~ R_{\rm exp} = 0.65$ and 
$\Delta\simeq 2\epsilon ~\mbox{for}~ R_{\rm exp} = 0.80$
The bound deteriorates quickly as $R_{\rm exp} \to 1$. In addition,
Eq.~(\ref{corrACP}) implies that similar effect could in principle
generate an ${\cal O}(10\%) $ CP asymmetry which is significantly larger than the 
bound $\mbox{${\cal A}^{\rm dir}_{\rm CP}$}(B^\pm\to\pi^\pm K) \sim 1 \%$. 
Therefore, in order to understand whether a large CP asymmetry signals 
New Physics, and whether it is possible to obtain a bound on $\gamma$, 
it is imperative to study FSI to obtain an order of magnitude estimate of the 
effect. 

\section{Final State Rescattering}
Final state interactions arise as a consequence of the
unitarity of the ${\cal S}$-matrix,
${\cal S}^\dagger {\cal S} = 1$, and involve the rescattering of
physical particles in the final state.
The ${\cal T}$-matrix, defined by ${\cal S} = 1 + i {\cal T}$,
obeys the optical theorem:
\begin{equation}
{\cal D}isc~{\cal T}_{B \rightarrow f} \equiv {1 \over 2i}
\left[ \langle f | {\cal T} | B \rangle -
\langle f | {\cal T}^\dagger | B \rangle \right]
= {1 \over 2} \sum_{i} \langle f | {\cal T}^\dagger | i \rangle
\langle i | {\cal T} | B \rangle \ \ ,
\label{unit}
\end{equation}
where ${\cal D}isc$ denotes essentially the imaginary part.
Using $CPT$ in the form
$
\langle \bar f | {\cal T} | \bar B \rangle^* =
\langle \bar B | {\cal T^\dagger} | \bar f \rangle =
\langle f | {\cal T^\dagger} | B \rangle
$
this can be tranformed into the
more intuititve form
\begin{equation}
\label{opt}
\langle \bar f | {\cal T} | \bar B \rangle^* =
\sum_{i}
\langle f | {\cal S}^\dagger | i \rangle
\langle i | {\cal T} | B \rangle \;\;.
\end{equation}
Here, the states $| i \rangle$ represent all possible final states
(including $| f \rangle$ itself)
which can be reached from the state  $| B \rangle$
by the weak transition matrix ${\cal T}$. The
right hand side of  Eq.~(\ref{opt})
can then be viewed as a weak decay of $|  B \rangle$
into $| i \rangle$ followed by a strong rescattering
of $| i \rangle$ into $| f\rangle$. Thus, we
identify $\langle f | {\cal S}^\dagger | i \rangle$
as a CP-conserving FSI rescattering of particles.
Notice that if $| i \rangle$ is an eigenstate of ${\cal S}$
with a phase $e^{2i\delta}$, we have
\begin{equation}
\label{triv}
\langle \bar i| {\cal T} | \bar B \rangle^* =
e^{-2i\delta_i}\langle i | {\cal T} | B \rangle \;\;.
\end{equation}
which implies equal rates for the charge conjugated
decays and hence no CP asymmtery. Therefore, at least
two different states with equal quantum numbers must exist
which can be connected by strong rescattering. Also
\begin{equation}
\langle \bar i | {\cal T} | \bar B \rangle = e^{i\delta} T_i
\langle i | {\cal T} | B \rangle = e^{i\delta} T_i^*
\label{watson}
\end{equation}
The matrix elements $T_i$ are assumed to be
the ``bare'' decay amplitudes, calculated e.g. in
factorization approximation~\cite{fact} and have
no rescattering phases. This implies that these transition
matrix elements between charge conjugated states
are just the complex conjugated ones of each other.
Eq.~(\ref{watson}) is known as Watson's theorem
\cite{watson52}. 

The above considerations allow to formulate a condition for 
the non-vanishing CP asymmetry. It requires two different
final states which undergo strong transitions into each 
other. The strong phase is then nothing but the occurence of the
{\it physical} intermediate state $| B_\beta \rangle$ and arises
when summing over the intermediate states.

The final state rescatterings of high energy particles
may be divided into `soft' and `hard' scattering.  Soft scattering
occurs primarily in the forward direction with limited transverse
momentum, having a distribution which falls exponentially 
with a scale of order $0.5$~GeV. Soft scattering might be best 
described by hadronic states. At higher transverse momentum
one encounters the region of hard scattering, which
falls only as a power of the transverse momentum.  Collisions
involving hard scattering are interpreted as interactions between
the quarks and gluons of QCD. Note that it is possible to generate FSI phases 
in nonleptonic B-decays into charmless final states in perturbative 
QCD~\cite{lw,bss,gerard}: there are two ways to reach a 
given final state, via the tree diagram $b \to u \bar u s$, 
and via $b \to c \bar c s$ process, with subsequent final state rescattering 
of the two charmed quarks into two up quarks (penguin diagram). Since the 
energy release in b-decay is of the order $m_b > 2m_c$, the rescattered c-quarks 
can go on-shell generating CP conserving phase and thus 
${\cal A}_{CP}^{dir}$.

For the soft FSI, the low energy effective theory of strong interactions 
can be used to reliably estimate FSI phase differences in the kaon system. 
In the $D$ system final state rescattering has been  
studied assuming the dominance of intermediate resonances~\cite{buccella}. 
In the $B$ system, where the density of the resonances 
available is large due to the increased energy, a different approach must be
employed. One can use the fact that the $b-$quark mass 
is large compared to the QCD scale. Then, the leading order behavior of 
soft final state phases in the $m_b \to \infty$ limit can be investigated. 
This can be done by considering first the {\it elastic} channel, and
demonstrating that elastic rescattering does not disappear
in the limit of large $m_B$. Since the unitarity of the ${\cal S}$-matrix
requires that the inelastic channels are indeed the dominant
contributors to soft rescattering, such contributions have to share
a similar behavior in the heavy quark limit.
The elastic channel is convenient because of the optical theorem
which connects the forward (imaginary) invariant amplitude 
${\cal M}$ to the total cross section,
\begin{equation}
{\cal I}m~{\cal M}_{f\to f} (s, ~t = 0) = 2 k
\sqrt{s} \sigma_{f \to {\rm all}} \sim s \sigma_{f \to {\rm all}} \ \ ,
\label{opti}
\end{equation}
where $s$ is the squared center-of-mass energy and $t$ is the squared
momentum transfer.
The asymptotic total cross sections are known
experimentally to increase slowly with energy and
can be parameterized by the form \cite{pdg,ld}:
\begin{equation}
\sigma (s) = X \left({s\over s_0}\right)^{0.08}
+ Y \left({s\over s_0}\right)^{-0.56} \ \ ,
\label{pl}
\end{equation}
where $s_0 = {\cal O}(1)$~GeV is a typical hadronic scale.
Thus, the imaginary part of the forward elastic scattering amplitude
(\ref{opti}) increases asymptotically as $s^{1.08}$.
Considering only the imaginary part of the amplitude, and building in 
the known exponential fall-off of the elastic cross section in $t$  
($t<0$)
\cite{jc} by writing
\begin{equation}
i{\cal I}m~{\cal M}_{f\to f} (s,t) \simeq i \beta_0 \left( {s \over s_0}
\right)^{1.08} e^{bt} \ \ ,
\label{fall}
\end{equation}
one can calculate the contribution
of the imaginary part of the elastic amplitude to the
unitarity relation for a final state $f = a + b$ with kinematics
$p_a' +  p_b' = p_a +  p_b$ and $s = (p_a + p_b )^2$:
\begin{eqnarray}
{\cal D}isc~{\cal M}_{B \to f} &=&
{1 \over 2} \int {d^3p_a' \over (2\pi)^3 2E_a'}
{d^3p_b' \over (2\pi)^3 2E_b'}
(2 \pi)^4 \delta^{(4)} (p_B - p_a' - p_b') \Big( -i\beta_0
\left( {s \over s_0} \right)^{1.08} e^{b(p_a - p_a')^2} \Big)
{\cal M}_{B \rightarrow f} \nonumber \\
&=& - {1\over 16\pi} {i\beta_0 \over s_0 b}\left( {m_B^2 \over s_0} 
\right)^{0.08} {\cal M}_{B \rightarrow f} \ \ ,
\label{mess}
\end{eqnarray}
where $t = (p_a - p_a')^2 \simeq -s(1 - \cos\theta)/2$,
and $s = m_B^2$.

On can refine the argument further, since
the phenomenology of high energy
scattering is well accounted for by the Regge theory~\cite{jc,zheng}.
In the Regge model, scattering amplitudes are described by the 
exchanges of Regge trajectories (families of particles of differing 
spin) with the leading contribution given by the Pomeron
exchange. Calculating the Pomeron contribution to the
elastic final state rescattering in $B \to \pi \pi$
one finds \cite{dgps}
\begin{equation}
{\cal D}isc~{\cal M}_{B \to \pi\pi}|_{\rm Pomeron} = -i\epsilon
{\cal M}_{B \to \pi\pi}, ~~~~~
\epsilon \simeq 0.21 \ \ .
\label{despite}
\end{equation}
It is important that the Pomeron-exchange amplitude is seen to be almost 
purely imaginary. However, of chief significance is the
identified weak dependence of $\epsilon$ on $m_B$  -- the
$(m_B^2)^{0.08}$ factor in the numerator is attenuated by the
$\ln(m_B^2/s_0)$ dependence in the effective value of $b$.

The analysis of the elastic channel suggests that, at high energies,  
FSI phases are {\it mainly generated by inelastic effects}.
This conclusion immediately follows from the fact that
the high energy cross section is mostly inelastic. This also
follows from the fact that the Pomeron elastic  
amplitude is almost purely imaginary.  Since the study of
elastic rescattering has yielded a ${\cal T}$-matrix element ${\cal  
T}_{ab
\to ab} = 2 i \epsilon$, i.e. ${\cal S}_{ab \to ab} = 1- 2
\epsilon$, and since the constraint of unitarity of
the ${\cal S}$-matrix
implies that the
off-diagonal elements are ${\cal O}(\sqrt{\epsilon})$,
with $\epsilon$
approximately ${\cal O}(m_B^0)$ in powers of $m_B$ and numerically
$\epsilon < 1$, then the inelastic amplitude must also be ${\cal
O}(m_B^0)$ and of magnitude $\sqrt{\epsilon} > \epsilon$.
Similar conclusions follow from the consideration of 
the final state unitarity relations.

The very presence of inelastic effects suggests a physical picture 
similar and complimentary to the ``color transparency argument''. 
This argument suggests that a
``small'' (compared to the typical hardonic size $1/\Lambda_{QCD}$) 
color-singlet two-quark configuration does not interact 
with the soft gluonic environment. Thus, if the two-body decay 
is dominated by this particular quark configuration with all other quark 
configurations yielding multiparticle final states, 
one might expect that the effects of FSI are suppressed in the decays of 
ultra-heavy particles. However, this quark configuration is realized only 
on the edge of the available phase space. Therfore, in the limit 
$m_b \to \infty$ one ultimately encounters the situation where the quarks 
easily combine to form a multiparticle state which then undergoes rescattering
into the two-body final state. Analysis of the final-state unitarity relations 
in their general form, 
\begin{equation}
{\cal D}isc~{\cal{M}}_{B \to f_1} = {1\over 2}\sum_k\ {\cal M}_{B\to
k} T^\dagger_{k \to f_1} \ \ ,
\label{unit2}
\end{equation}
is complicated due to the many contributing intermediate states 
present at the $B$ mass.  However, it is possible to illustrate 
the systematics of inelastic scattering in a simple 
two-channel model.  This example involves a two-body final state 
$f_1$ undergoing elastic scattering and a final state $f_2$ which 
represents `everything else'.  We assume that the elastic amplitude 
is purely imaginary.  Thus, the scattering can be described 
in the one-parameter form 
\begin{equation}
 S = \left( \begin{array}{cc} 
\cos 2 \theta & i \sin 2 \theta \\
              i \sin 2 \theta & \cos 2 \theta
\end{array} \right)  \ ,\qquad \qquad 
 T = \left( \begin{array}{cc}
2 i \sin^2 \theta &  \sin 2 \theta \\
               \sin 2 \theta & 2 i \sin^2 \theta 
\end{array} \right)  \ \ ,
\label{matr1}
\end{equation}
where, from our elastic-rescattering calculation, we identify $\sin^2 
\theta \equiv \epsilon$. The unitarity relations become
\begin{eqnarray}
{\cal D}isc ~{\cal{M}}_{B \to f_1} &=& 
- i \sin^2 \theta {\cal{M}}_{B \to f_1} +
\frac{1}{2} \sin 2 \theta {\cal{M}}_{B \to f_2} \ \ ,\nonumber \\
{\cal D}isc~ {\cal {M}}_{B \to f_2} &=& \frac{1}{2} \sin 2 \theta 
{\cal{M}}_{B \to f_1} - i \sin^2 \theta {\cal{M}}_{B \to f_2} \ \ 
\label{big}
\end{eqnarray}
Denoting ${\cal{M}}_1^0$ and ${\cal{M}}_2^0$ 
to be the decay amplitudes in the limit $\theta \to 0$,
an exact solution to Eq.~(\ref{big}) is given by
\begin{equation}
{\cal{M}}_{B \to f_1} = \cos \theta {\cal{M}}_1^0 + i \sin \theta
{\cal{M}}_2^0  \ , \qquad 
{\cal{M}}_{B \to f_2} = \cos \theta {\cal{M}}_2^0 + i \sin \theta
{\cal{M}}_1^0 \ \ .
\label{soln}
\end{equation}
In this example we see that the phase is given by the inelastic
scattering with a result of order 
\begin{equation}
\frac{ {\cal I}m~ {\cal{M}}_{B \to f}}{{\cal R}e~ {\cal{M}}_{B \to f}} \sim 
\sqrt{ \epsilon}~ \frac{{\cal{M}}_2^0}{{\cal{M}}_1^0} \ \ .
\end{equation}
Clearly, for physical $B$ decay, we no longer 
have a simple one-parameter ${\cal S}$ matrix ,and,
with many channels, cancellations or enhancements are 
possible for the sum of many contributions.
However, the main feature of the above result is expected 
to remain -- that inelastic channels cannot vanish because they 
are required to make the discontinuity real and that 
the phase is systematically of order $\sqrt{\epsilon}$ from these
channels and thus does not vanish in the large $m_B$ limit. Moreover, 
it is possible to show that inelastic FSI can contribute to CP violating
asymmetries at the leading order in $m_B$~\cite{dgps}. Non-zero 
FSI phases have been recently observed by CLEO~\cite{CLEOfsi}.

We should note that radiative weak decays of B mesons
are also affected by FSI. For instance, the
extraction of $V_{td}$ from $B \to \rho \gamma$ is hampered by  
uncertainties related to certain long distance effects  
from the on-shell hadron rescattering with subsequent conversion 
of one of the hadrons into the photon. This contribution could be 
sizeable~\cite{dgp}.

\section{Testing Models of Final State Interactions}

{\it (i) Model Independent Bounds on the FSI Corrections}.
In view of the large theoretical uncertainties involved in the calculation
of the FSI contributions, it would be extremely useful to find a phenomenological 
method by which to bound the magnitude of the FSI contribution. The observation 
of a larger asymmetry would then be a signal for New Physics. 
Here the application of  flavor $SU(3)$ flavor symmetry, provides powerful
methods to obtain a direct upper bound on the FSI contribution. 

The simplest example involves bounding FSI in $B \to \pi K$ decays 
using $B^\pm \to K^\pm K$ transitions.
The effective Hamiltonian for $b \to d$ decays may be obtained from
(\ref{HeffCP}) by the substitution $s \to d$. In analogy with 
Eq.~(\ref{Camps}) the amplitudes may be decomposed according to 
their dependence on CKM factors, giving
\begin{eqnarray} \label{KKamps}
A(B^+ \to K^+ \bar K^0 ) &=& A_{cd} - A_{ud} e^{i \gamma} e^{i \delta}\,,\nonumber\\
A(B^- \to K^- K^0 ) &=& A_{cd} - A_{ud} e^{- i \gamma} e^{i \delta}\,. 
\end{eqnarray}
Invariance under the $SU(3)$ rotation
$\exp(i {\pi \over 2} \lambda_7 )$, i.e., interchange of $s$ and $d$ quark fields,
implies equalities among operator matrix elements and amplitudes,
\begin{eqnarray} \label{su3} 
\langle K^- K^0 | Q_i^{qd} | B^- \rangle = 
\langle K^0 \pi^- | Q_i^{qs} | B^- \rangle,&&~~~q=u,c,;~~i = 1,2 \nonumber \\ 
\langle K^- K^0 | Q_i^{d} | B^- \rangle = 
\langle K^0 \pi^- | Q_i^{s} | B^- \rangle,&& ~~~i = 3,..,10\,. \\
A_{ud} e^{i \delta} = A_{us}^+ e^{i \delta_+}
{V_{ud} \over V_{us} } (1 + R_{ud} )\,,&&
~~~A_{cd} = A_{cs}^+ {V_{cd} \over V_{cs} } (1 + R_{cd} ), \nonumber
\end{eqnarray}
where $R_{ud}$ and $R_{cd}$ parameterize $SU(3)$ violation, which
is typically of the order of $20-30\%$.  Note that it is only an $SU(2)$ subgroup
of $SU(3)$, namely $U$-spin, which is required to derive these relations.  Since
the $B^-$ carries $U=0$ and the transition operators $Q_i^{qd}$ and $Q_i^d$
carry $U={1\over2}$, it is only the $U={1\over2}$ component of the $K^-K^0$
final state which couples to the decay channel.  As a result, 
an upper bound on $\epsilon$ follows from the ratio
\begin{equation} \label{RK} R_K = {BR(B^+ \to K^+ \overline{K^0} ) +
BR(B^- \to K^- K^0 ) \over BR(B^+ \to K^0
\pi^+ ) +BR(B^- \to \overline{K^0}  \pi^- )}\,. 
\end{equation}
After some algebra, we obtain for $\epsilon$ and ${\cal A}^{\rm dir}_{\rm CP}$
\begin{eqnarray} \label{rmax}
\epsilon < \lambda \sqrt{R_{K}}\, (1 + {\rm Re}[R_{ud} ]) + \lambda^2 (R_{K} +
1 ) \cos\gamma \cos \delta_+  + {\cal O}(\lambda^3 ,
\lambda^2 R_{ud,cd} )\,, \nonumber \\
 |\mbox{${\cal A}^{\rm dir}_{\rm CP}$}| < 2 \lambda \sqrt{R_{K} R}\, (1 + Re[R_{ud}])+ 2
\lambda^2 \sqrt{R}\, (R_K\sqrt{1-R}+R_{K} + 1)
+ {\cal O}(\lambda^3 , \lambda^2 R_{ud,cd} )\,.
\end{eqnarray}
Using the experimental bound $R_K < 1.9 $ one obtains
$\epsilon < 0.4 $ and $\mbox{${\cal A}^{\rm dir}_{\rm CP}$} < 0.6 $.
More interesting constraints on $\epsilon$ and $\mbox{${\cal A}^{\rm dir}_{\rm CP}$}$ 
could also be be obtained \cite{bkp}. 

{\it (ii) Direct Observation}. As we have shown above, the amplitude for the decay  
of a $B$-meson into some final state $f$ includes a direct contribution 
$A(B \rightarrow f)$ and a sum over the contributions 
$A(B \rightarrow i \rightarrow f)$, which corresponds to the weak decays of the 
$B$-meson into intermediate hadronic states $i$, followed by the strong 
scattering of $i$ into $f$. The possibility of significant final state 
scattering effects becomes real when considering rare decays, for 
which the amplitude $A(B \rightarrow f)$ is suppressed compared to 
$A(B \rightarrow i)$. In the ideal case, when the direct contribution
is absent, one may be able to isolate the effect of FSI completely.
While this situation might not be realized in the nature, rare weak decays
offer tantalizing possibility of the {\it direct observation} of the effects
of FSI. 

One of the possibilities involve dynamically suppressed decays which 
proceed via weak annihillation diagrams. It has been argued that
final state interactions can modify the decay amplitudes,
violating the expected hierarchy of amplitues. For example, it is 
expected that the amplitudes that do not
involve spectator quarks (such as tree-level or penguin amplitudes)
dominate over the diagrams involving spectator quarks
(e.g. weak annihilation or weak rescattering amplitudes).
In many cases, large amplitudes might contribute to
the processes involving spectator quarks through the final
state rescattering~\cite{bh,bgr}. It must be stressed
that although the predictive power of available methods is limited and
most of the estimates are based on the two-body
rescattering diagrams, some conclusions can still be reached.
For instance, it is possible to show \cite{bgr}
that the rescattering from the dominant channel leads to the
suppression only of the order $\lambda \sim 0.2$ compared to
$f_B/m_B \sim \lambda^2$ obtained from the naive quark
diagram estimate. 

Another class of decays involve the so-called
OZI-violating modes, i.e. those which cannot be realized in 
terms of quark diagrams without annihillation of 
at least one pair of the quarks. It includes modes like
$\overline{B}^0_d \rightarrow \phi \phi, D^0 \phi$ and 
$J/\psi \phi$. For instance, the direct contribution 
to $\overline{B}^0_d \to \phi \phi$ involves a suppressed 
space-like penguin diagram. However, the unitarity of the
${\cal S}$-matrix, ${\cal S}^\dagger {\cal S} = 1$, implies
that this decay can also proceed via the OZI-allowed weak
transition followed by final state rescattering into the
final state under consideration. These 
two-step OZI violating processes were intensively studied
in connection with certain low-energy processes
\cite{GeIsgur,lipkin}. In $B$-decays these OZI-allowed
steps involve multiparticle intermediate states and
might provide a source for significant violation of 
the OZI rule. For instance, the FSI contribution can proceed
via $\overline{B}^0_d \rightarrow \eta^{(\prime)} 
\eta^{(\prime)} \rightarrow \phi \phi$, $\overline{B}^0_d \rightarrow 
D^{\ast 0} \eta^{(\prime)} \rightarrow D^0 \phi$ and $\overline{B}^0_d 
\rightarrow  \psi' \eta^{(\prime)} \rightarrow J/\psi \phi$.
The intermediate state might also include an additional set
of pions. The weak decay into the intermediate state occurs at tree level, 
through the $(u\overline{u} + d\overline{d})/\sqrt{2}$ component of the 
$\eta^{(\prime)}$ wavefunction, whereas the strong scattering into the 
final state involves the $s\bar{s}$ component~\cite{donoghue}.
Hence the possibility of using these decay modes as direct probes of the 
FSI contributions to $B$ decay amplitudes. It is however possible to show that
there exist strong cancellations\cite{lipkin,dgps2} among various 
{\it two body} intermediate channels. In the example of 
$\overline{B}^0_d \to \phi \phi$, the cancellation among
$\eta$ and $\eta'$ is almost complete, so the
effect is of the second order in the $SU(3)$-breaking corrections
\begin{equation}
Disc~{\cal M}_{B \to \phi \phi} = O(\delta^2, \Delta^2, 
\delta \Delta) f_\eta F_0 A, ~~\delta = f_{\eta'} - f_\eta,~
\Delta = F_0' - F_0,
\end{equation}
with $A \sim s^{\alpha_0 - 1} e^{i\pi \alpha_0 /2} / 8 b$.
This implies that the OZI-suppressed decays provide an excellent
probe of the multiparticle FSI. Given the very clear signature, these 
decay modes could be probed at the upcoming $B$-factories.






\end{document}